\documentstyle[11pt,mynewpasp,twoside,epsf]{article}
\def\ML{\mbox{$M/L$}}

\def\edcomment#1{\iffalse\marginpar{\raggedright\sl#1\/}\else\relax\fi}
\reversemarginpar

\begin{document}
\title{Is there a dichotomy in the Dark Matter as well as in the Baryonic Matter properties of ellipticals?}
 \author{N.R. Napolitano$^1$, M. Capaccioli$^2$, M. Arnaboldi$^3$, M.R. Merrifield$^4$, N.G. Douglas$^1$, K. Kuijken$^{1,5}$, A.J. Romanowsky$^4$, K.C. Freeman$^6$}
\affil{$^1$ Kapteyn Institute, Groningen; $^2$ INAF-Osservatorio Astronomico di Capodimonte, Naples; $^3$ INAF- Osservatorio Astronomico di Pino Torinese, Turin; $^4$ School of Physics \& Astronomy, University of Nottingham; $^5$ University of Leiden; $^6$ RSAA, Mt. Stromlo Observatory}

\begin{abstract}
We have found a correlation betwen the \ML\ global gradients and the structural
parameters of the luminous components of a sample of 19 early-type galaxies. Such a
correlation supports the hypothesis that there is a connection between
the dark matter content and the evolution of the baryonic component in
such systems.
\end{abstract}

\section{Background and new evidence}

There are several lines of evidence for a dichotomy in the properties of
early-type galaxies: fainter systems have pointed (disky) isophotes and central power-law surface brightness profiles, while bright galaxies are boxy and show central cores (Nieto \& Bender 1989, Faber et al. 1997). 
This dichotomy has been interpreted in an evolutionary framework: disky/faint systems have not experienced merger events in the recent past (Nieto \& Bender 1989), or alternatively are remnants of gas-rich merging events (Faber et al. 1997), while bright/boxy systems are probable merger remnants (Nieto \& Bender 1989, Faber et al. 1997).
This scheme is supported by X-ray properties of early-types (Pellegrini (1999) showed that faint/disky/power-law early-type galaxies are also fainter in X-ray luminosity, while bright/boxy/core galaxies are X-ray bright) and GCs number densities (Kissler-Patig 1997). What then is the actual mechanism which has triggered the evolution of both the stellar and hot gas components in galaxies? 

In Fig. 1 we plot the global M/L radial gradients, $\Delta  \Gamma/\Delta  \mathrm{R}$  ($\Gamma=M/L_B$), based on planetary nebulae kinematics and long-slit spectroscopy archive data, as a function of the intrinsic absolute magnitude, the isophotal shape parameter $a_4$, and the
$\gamma$ parameter, i.e. the slope of the surface brightness
profile in the galaxy core ($\sim R^{-\gamma}$).
Fig. 1 suggests a general regularity of the \ML\ gradients with respect to the structural parameters for the majority of the galaxies in the sample, except for a few cases (open symbols): these are noted in literature as interacting candidates since they show dynamical peculiarities suggesting they are not in
equilibrium.
If we exclude this subsample, with very steep ``apparent'' \ML\ gradients, we see that smaller gradients ($\Delta \Gamma_B/\Delta R \le
0.8$) are found for systems with faint total
magnitudes ($M_B>-20$), mostly disky ($100\times a_4/a>0.2$) and
power-law ($\gamma>0.15$), while bright/boxy/core galaxies show larger
gradients ($0.8<\Delta \Gamma_B/\Delta R \le 2.7$). We have found these trends significant at better than 95\% c.l. via the Spearman Rank test.


\begin{figure}[t]  
\vspace{-6cm}
\plotone{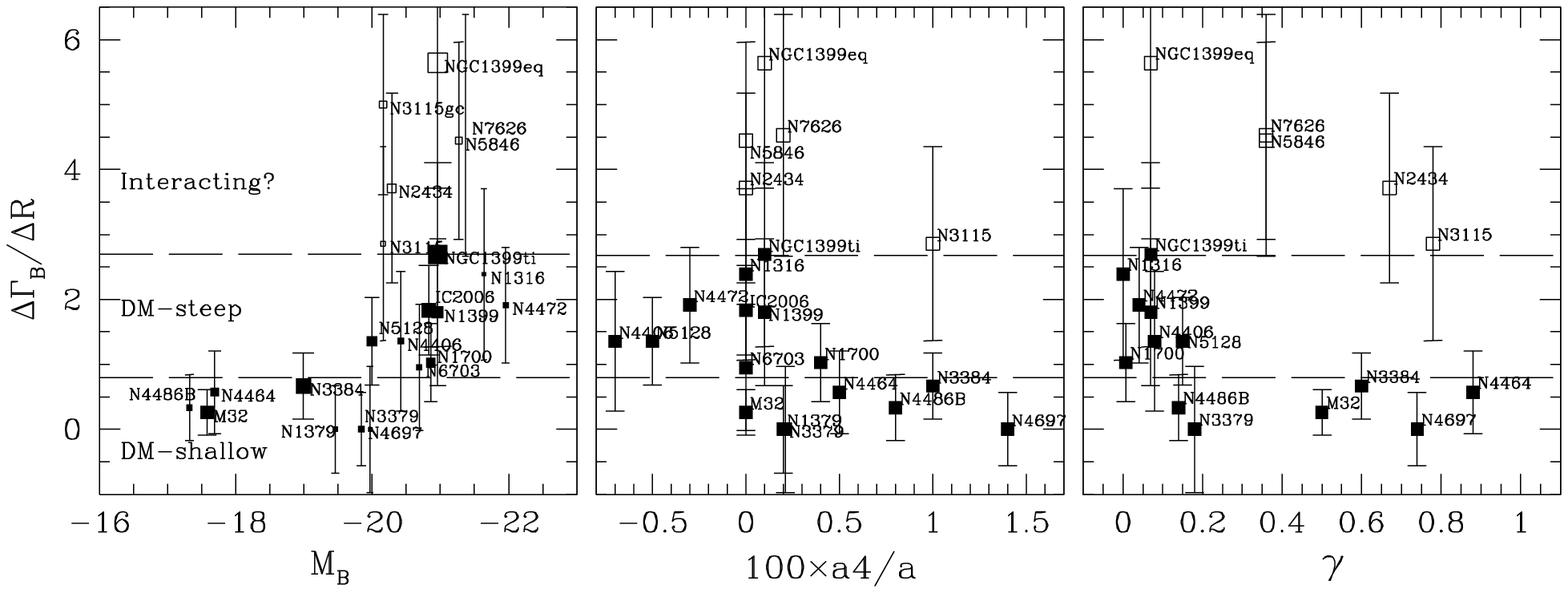}
\vspace{-2.5cm}
\caption{\small $M/L$ gradients as a function of the intrinsic
parameters. Dashed lines show a tentative separation between shallow and steep gradients and the interacting sample (see discussion in the text). In the left panel symbol dimensions are
proportional to the outermost radii where \ML\ estimates are
available.}
\end{figure}

\section{Discussion and Conclusions}
Following an earlier suggestion (Capaccioli et al. 2002), we have found that the dichotomy in the early-type galaxies with respect to their structural parameters and X-ray properties, see Pellegrini 1999) seems to correspond to a trend in the \ML\ global radial gradients. 
This possibly indicates that the dark matter has triggered the evolution of
both the stellar and hot gas components in galaxies. For instance, Eskridge et al. (1995) and Matsushita (2001)
have suggested that the correlation of the hot gas assembly in
early-type galaxies and the depth of the potential wells could explain the correlation of the X-ray luminosity with the shape and dynamical
parameters ($a_4$, axial ratio and central velocity dispersion). Here we can confirm that faint/disky galaxies do, indeed, have shallower potential wells when compared to bright/boxy galaxies.

\end{document}